\documentclass[a4paper,english,aps,twocolumn]{revtex4}
\usepackage[T1]{fontenc}
\usepackage[latin9]{inputenc}
\usepackage{graphicx}

\makeatletter

\usepackage{textcomp}

\usepackage{esint}

\makeatletter
\@ifundefined{textcolor}{}
{%
 \definecolor{BLACK}{gray}{0}
 \definecolor{WHITE}{gray}{1}
 \definecolor{RED}{rgb}{1,0,0}
 \definecolor{GREEN}{rgb}{0,1,0}
 \definecolor{BLUE}{rgb}{0,0,1}
 \definecolor{CYAN}{cmyk}{1,0,0,0}
 \definecolor{MAGENTA}{cmyk}{0,1,0,0}
 \definecolor{YELLOW}{cmyk}{0,0,1,0}
 }

\makeatother

\usepackage{babel}

\makeatother

\usepackage{babel}

\begin{document}

\title{Exchange bias in $Co$/$CoO$ core-shell nanowires: Role of the antiferromagnetic
superparamagnetic fluctuations}

\author{Thomas Maurer, Fatih Zighem, Frédéric Ott\footnote{Frederic.Ott@cea.fr}, Grégory Chaboussant, Gilles
André}

\affiliation{CEA, IRAMIS, Laboratoire Léon Brillouin, F-91191 Gif sur Yvette,
France }

\affiliation{CNRS, IRAMIS, Laboratoire Léon Brillouin, F-91191 Gif sur Yvette,
France.}

\author{Yaghoub Soumare and Jean-Yves Piquemal}

\affiliation{ITODYS Université Paris 7 - Denis Diderot, UMR CNRS 7086, 15 rue
Jean-Antoine de Baïf F-75205 Cedex 13 Paris, France}

\author{Guillaume Viau}

\affiliation{Université de Toulouse, LPCNO, INSA, UMR CNRS 5215, 135 avenue de
Rangueil, F-31077 Toulouse Cedex 4.}

\author{Christophe Gatel}

\affiliation{Université de Toulouse, CEMES, UPR CNRS 8011, 29 rue Jeanne Marvig,
BP94347, F-31055 Toulouse Cedex, France}
\begin{abstract}
The magnetic properties of Co ($<D>=15$ {\normalsize nm}, $<L>=130$
{\normalsize nm}) nanowires are reported. In oxidized wires, we measure
large exchange bias fields of the order of $0.1$ {\normalsize T}
below $T\sim100$ {\normalsize K}. The onset of the exchange bias,
between the ferromagnetic core and the anti-ferromagnetic CoO shell,
is accompanied by a coercivity drop of $0.2\, T$ which leads to a
minimum in coercivity at $\sim100$ K. Magnetization relaxation measurements
show a temperature dependence of the magnetic viscosity $S$ which
is consistent with a volume distribution of the CoO grains at the
surface. We propose that the superparamagnetic fluctuations of the
anti-ferromagnetic CoO shell play a key role in the flipping of the
nanowire magnetization and explain the coercivity drop. This is supported
by micromagnetic simulations. This behavior is specific to the geometry
of a 1D system which possesses a large shape anisotropy and was not
previously observed in $0D$ (spheres) or $2D$ (thin films) systems
which have a high degree of symmetry and low coercivities. This study
underlines the importance of the AFM super-paramagnetic fluctuations
in the exchange bias mechanism. 
\end{abstract}
\maketitle

\section{introduction}

Magnetic nanowires are of prime interest both scientifically and for
applications in the nanotechnology industry (in magnetic memories
\cite{parkin2008}, in magnetic recording media \cite{gapin2006},
in sensors \cite{McGary2006} or in microwave devices \cite{Ye2007}).
The magnetic properties of nanowires are essentially governed by the
very strong shape anisotropy giving rise to high coercive fields which
may have applications for permanent magnets fabrication \cite{Maurer2007}.
Exchange-biased systems like FM/AFM layers or core/shell FM/AFM spherical
particles are characterized by the Néel temperature $T_{N}$, corresponding
to the ordering of the antiferromagnetic layer, and the blocking temperature
$T_{EB}$ corresponding to the apparition of the exchange bias field
$H_{EB}$, usually lower than $T_{N}$ \cite{nogues1999,nogues2005,berkowitz1999,vanDerZaag2000-1,Hong2006-1}.
Most of the recent studies of the exchange bias mechanism have been
performed on thin film systems \cite{berkowitz1999,nogues1999,nogues2005,radu2008}
since they permit a good control of the thickness and textures and
the temperature dependence of the exchange field $H_{EB}$ has been
extensively studied \cite{nogues1999,Stamps}. On the other hand,
the temperature dependence of the coercivity is studied although it
can exhibit a variety of behaviors depending on the anisotropy of
the AFM layer \cite{nogues1999}. Actually, it can be difficult to
study the temperature dependence of the coercivity due to the microstructured
character of the material \cite{nogues1999}. However, it has been
shown that, in the case of exchange-biased systems whose AFM layer
exhibits a small anisotropy, a coercivity peak can arise around the
blocking temperature \cite{nogues1999}. In this article, we discuss
the exchange bias properties of Co nanowires with large coercive fields,
mainly due to the 1D geometry \cite{Maurer2007} , and show that this
1D character leads to a specific exchange bias behavior, in connection
with the superparamagnetic relaxation of the CoO grains present at
the surface of the nanowires. The paper is organized as follows: in
Section II, we present the magnetic nanowires. Section III gives the
experimental results obtained from magnetometry measurements. The
experimental results are discussed in Section IV. Micromagnetic simulations
are presented in Section V.

\section{Magnetic nanowires}

Co nanowires \cite{Soumare2009} have been synthesized by reduction
in liquid polyol \cite{Ung2005,Soumare2008}. Transmission Electron
Microscopy (TEM) shows Co nanowires with a mean diameter $<D>$ of
$15$ nm and a mean length $<L>$ of $130$ nm (see Fig. 1). The standard
deviation on the diameter distribution $\sigma_{d}$ is small $(\sim10\%)$.
The length distribution is broader with a standard deviation $\sigma_{L}\sim20\%$.
The nanowires are well preserved from oxidation as long as they remain
in their polyol solution. In order to perform magnetic characterizations,
the nanowires are collected by centrifugation and washed several times
with ethanol. In this case, the wires oxidize at their surface \cite{Ung2005,Soumare2008}.
After a few weeks the system reaches a stable magnetic state via a
passivation mechanism \cite{gango1993,galland2005}.

\begin{figure}
\includegraphics[bb=30bp 69bp 341bp 483bp,clip,angle=270,width=7.5cm]{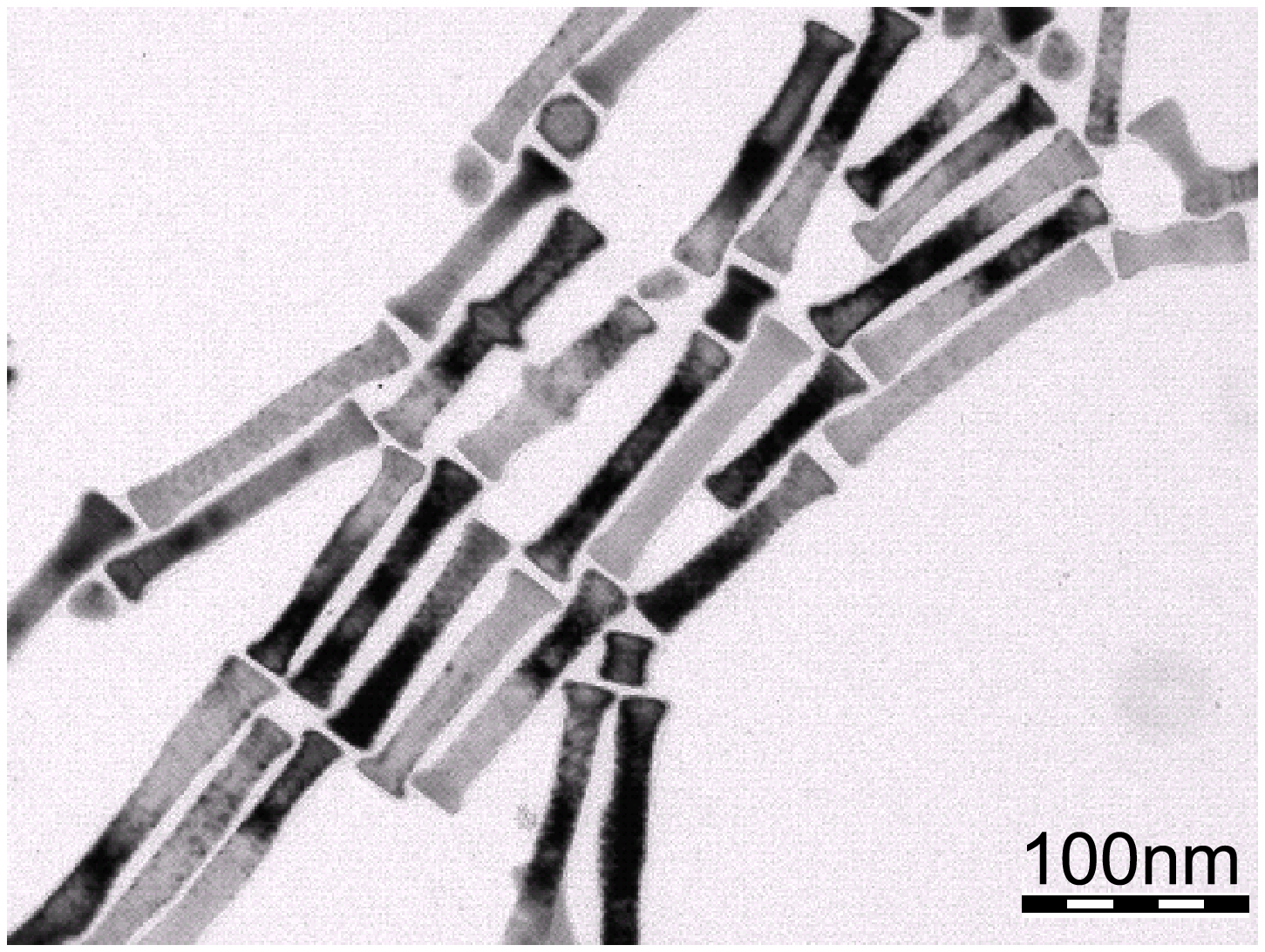}

\label{fig:1}

\caption{TEM image of $Co$ nanowires\cite{Soumare2009}. }

\end{figure}

High-resolution TEM (HRTEM) and X-ray diffraction show very well crystallized
wires in the metallic hcp phase with the crystallographic c-axis parallel
to the wires axis (see Fig. 2). The HRTEM image presented on Fig.
2 shows a wire with a mean diameter of 13 nm that consists of a core
of metallic cobalt coated by a thin oxide layer of CoO. The diffraction
pattern calculated from the image of the Co core was indexed as the
$[11\overline{2}0]$ zone axis of the hcp structure showing that the
c-axis is parallel to the wire axis. The core is nearly single crystal,
only few stacking faults diffuse lines are observed perpendicular
to the $[0002]$ direction. The CoO oxide layer is continuous all
over the wire edges. Its thickness inferred from HRTEM images is estimated
to $1.2\pm0.1$ nm on the edge of the wires and to $1.4\pm0.1$ nm
on the tips. Diffraction patterns calculated on the edge and on the
tip of the wire are indexed as the $[\overline{1}10]$ zone axis of
the fcc structure with two distances of 0.212 nm and four distances
of 0.245 nm corresponding respectively to the (002) and (111) reflections
of the \emph{Fm3m} cubic cobalt oxide CoO \cite{JCPDS}. The crystallographic
orientation relationships between the native oxide and the metal are:
CoO $[\overline{1}10]$ (111) // Co $[11\overline{2}0]$ (0001) and
CoO $[\overline{1}10]$(110) // Co $[11\overline{2}0]$ $(1\overline{1}00)$
on the tip and the edges, respectively. These relationships allow
to (i) minimize the mismatch between cobalt oxide and cobalt parameters
on the edges: 0.212 nm and 0.202 nm for the (200) oxide and (0002)
cobalt distances, respectively, and (ii) retain the hexagonal symmetry
of the hcp Co (0001) plane in the CoO (111) plane on the tip. The
oxide layer appears monocrystalline both on the tip and on the edges
but is globally polycrystalline because of the different orientations
on the wire facets. Therefore, from the bulk measurement point of
view, the CoO layer will be considered as disordered and composed
of crystallites of various sizes. 

The roughness of the interface between the Co core and the oxide layer
is smaller than $0.5$ nm showing that we have very well defined interfaces,
which have a quality equivalent to thin films deposited by vacuum
techniques. 

\begin{figure}
\includegraphics[bb=36bp 86bp 371bp 416bp,clip,angle=270,width=8cm]{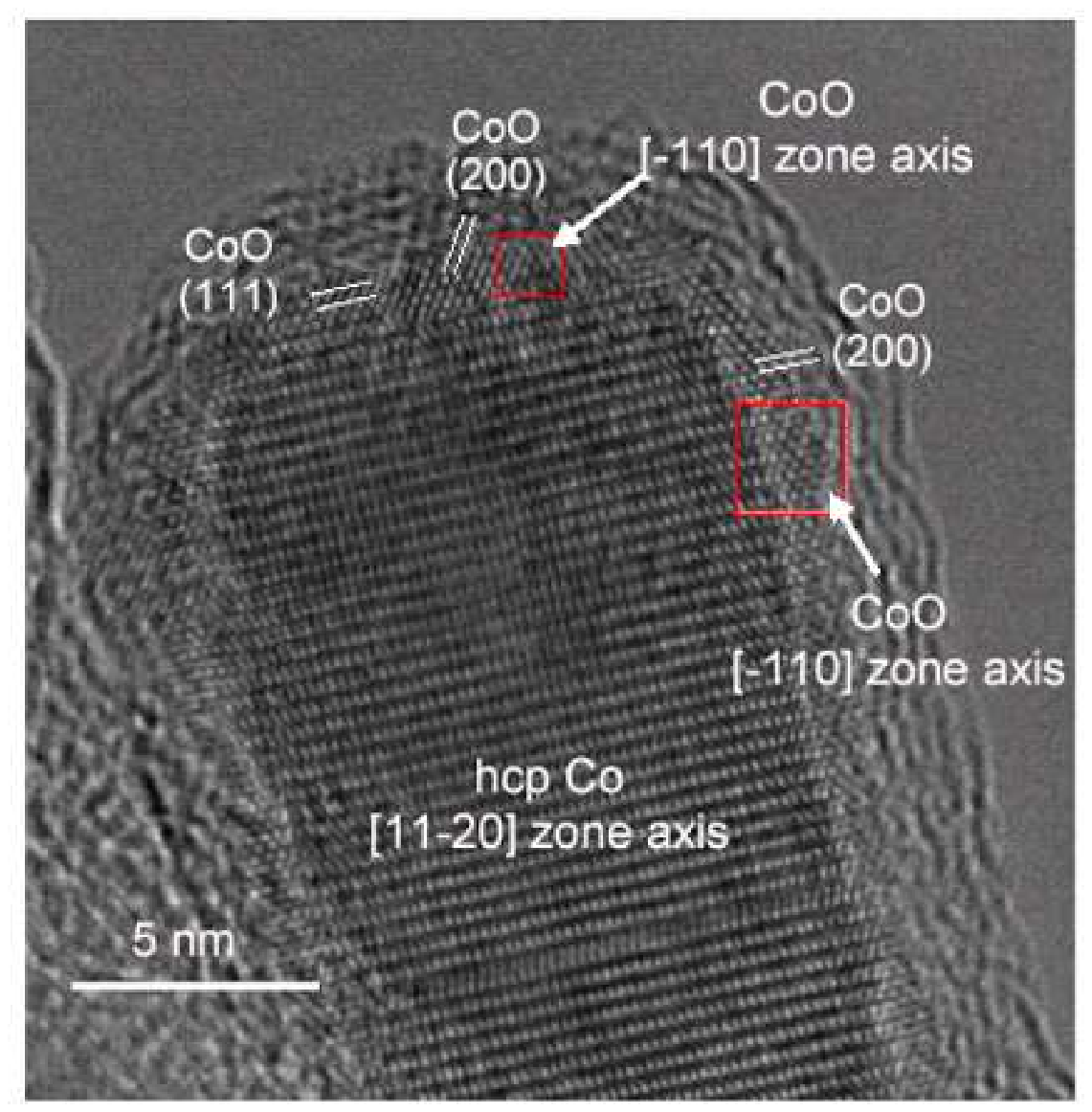}

\caption{HRTEM image of the tip of a Co wire showing the local structure of
the Co wire surrounded by a CoO shell (1.2 nm thickness).}

\end{figure}

Bulk CoO is an antiferromagnet (AFM) with a Néel temperature $T_{N}=293$
K. It has been shown that CoO layers as thin as $1$ nm on oxidized
Co particles still present AFM order close to room temperature \cite{gango1993,skumryev2003}
and that the Néel temperature in very thin epitaxial CoO layers can
even be increased well above room temperature \cite{vanDerZaag2000-1}.
In the case of Co/CoO nanospheres, a $T_{N}$ of about $235$ K was
reported \cite{inderhees2008}.

\begin{figure}
\includegraphics[bb=0bp 40bp 600bp 800bp,clip,width=8cm]{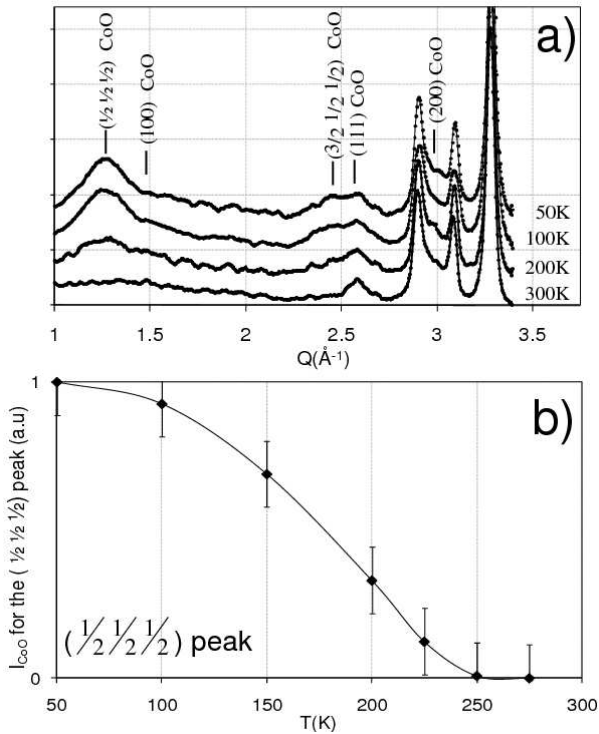}

\caption{(a) Neutron diffraction pattern for oxidized Co nanowires for temperatures
between 50K and 300K. The CoO peaks are indexed following the cubic
lattice diffraction pattern. The three non-indexed peaks above $Q=2.89\:\mathring{A}^{-1}$
correspond to the Co structural diffraction peaks \cite{Soumare2009}.
(b) Intensity of the $\left(\frac{1}{2}\,\frac{1}{2}\,\frac{1}{2}\right)$
CoO diffraction peak as a function of the temperature. The CoO shell
orders anti-ferromagnetically between $220$ K and $250$ K.}

\end{figure}

We performed neutron powder diffraction experiments on the G4.1 spectrometer
at the Laboratoire Léon Brillouin in order to determine $T_{N}$ (see
Fig. 3a). Above $T_{N}$, bulk CoO has the rocksalt structure \cite{shull1951}
whereas below $T_{N}$ there is a small trigonal and tetragonal distorsion
\cite{inderhees2008,tomiyasu2004,Jauch2001}. Thus bulk CoO crystal
structure becomes monoclinic (C2/m phase) when the antiferromagnetic
order sets in. However we indexed the peaks following the cubic lattice
diffraction pattern in a first approximation as it is usually done.
At room temperature we observe the two nuclear peaks (111) and (200)
at respectively $2.54\:\mathring{A}^{-1}$ and $2.94\:\mathring{A}^{-1}$.
When the temperature is lowered, three magnetic peaks appear: the
$\left(\frac{1}{2}\,\frac{1}{2}\,\frac{1}{2}\right)$ and $\left(\frac{3}{2}\,\frac{1}{2}\,\frac{1}{2}\right)$
peaks at respectively $q_{2,a}=1.27\:\mathring{A}^{-1}$ and $q_{2,b}=2.43\:\mathring{A}^{-1}$
from the AFM-II order and the (100) peak at $q_{1}=1.47\:\mathring{A}^{-1}$
from the AFM-I order. The temperature dependence of the AFM-II peaks
intensity (see Fig. 3b) shows that the AFM order sets in around $230$
K. This is comparable to what has been observed in Co/CoO spherical
particles \cite{inderhees2008}. The (100) peak of the AFM-I order
is barely visible. However gaussian fits of the pattern suggest that
the (100) peak appears only below 150K, contrary to what was observed
in \cite{inderhees2008}. Also note that above $250$ K, a very broad
magnetic diffuse scattering is observed around the $\left(\frac{1}{2}\,\frac{1}{2}\,\frac{1}{2}\right)$
position suggesting that AF correlations already exist at higher temperatures.
From the present data, we consider that the Néel temperature of the
$CoO$ shell around the wires is around $T_{N}=230$ K, which is lower
than the bulk value. Using the Scherrer formula, the width of the
$\left(\frac{1}{2}\,\frac{1}{2}\,\frac{1}{2}\right)$ peak corresponds
to a magnetic correlation length of $1-2$ nm. This is in agreement
with the thickness of the oxide shell. Note that neutron diffraction
measures an instantaneous picture of the AF ordering of the CoO shell
so that it is not sensitive to super-paramagnetic fluctuations (slower
than $10^{-14}$ s) of the small CoO crystallites. The measured $T_{N}$
temperature thus does not correspond to the blocking temperature of
the CoO crystallites.

\section{Experimental}

The nanowire powders were characterized by SQUID magnetometry. We
considered 2 types of samples: (i) non oxidized Co wires, which were
used as reference samples and kept in their butane-diol synthesis
solution \cite{viau_rep} and (ii) Co dried powders exposed to air,
which led to a natural oxidation. Fig. 4 presents the evolutions of
the exchange bias field $H_{EB}$ and coercive field $H_{C}$ for
these systems as a function of temperature. In these measurements,
the samples were field-cooled under $5$ T and the hysteresis cycles
were measured while increasing the temperature from $5$ K to $300$
K. The hysteresis cycles have been obtained by applying external magnetic
fields ranging from $-5$ T to $+5$ T. The hysteresis loops are not
perfectly square, and the magnetization at remanence is around 0.7$M_{s}$
for non oxidized wires and around 0.6$M_{s}$ for oxidized wires.
Two typical hysteresis cycles are presented in Fig. 4a.

In the case of non oxidized samples (triangles in Fig. 4b), no exchange
bias is observed and the coercive field decreases monotonously from
$\mu_{0}H_{c}\sim0.9$ T at low temperatures to $0.5$ T at room temperature.
In the case of oxidized samples (circles), an exchange bias field
$H_{EB}$ appears below $T_{EB}\approx100$ K. This exchange bias
field reaches $0.1$ T at low temperatures. The most striking feature
is that the coercive field $H_{C}$ dependence is not monotonous since
the coercive field decreases down to a \emph{minimum} at $T_{EB}$,
then reaches a maximum at about $200$ K, and finally decreases again
when reaching room temperature. It thus appears that $\mu_{0}H_{C}$
is maximum ($0.6\, T$) at $T\approx T_{N}\approx200$ K. Similar
results (not shown here) were obtained on 3 different batches of samples.

\begin{figure}
\includegraphics[bb=42bp 74bp 414bp 320bp,clip,angle=270,width=8cm]{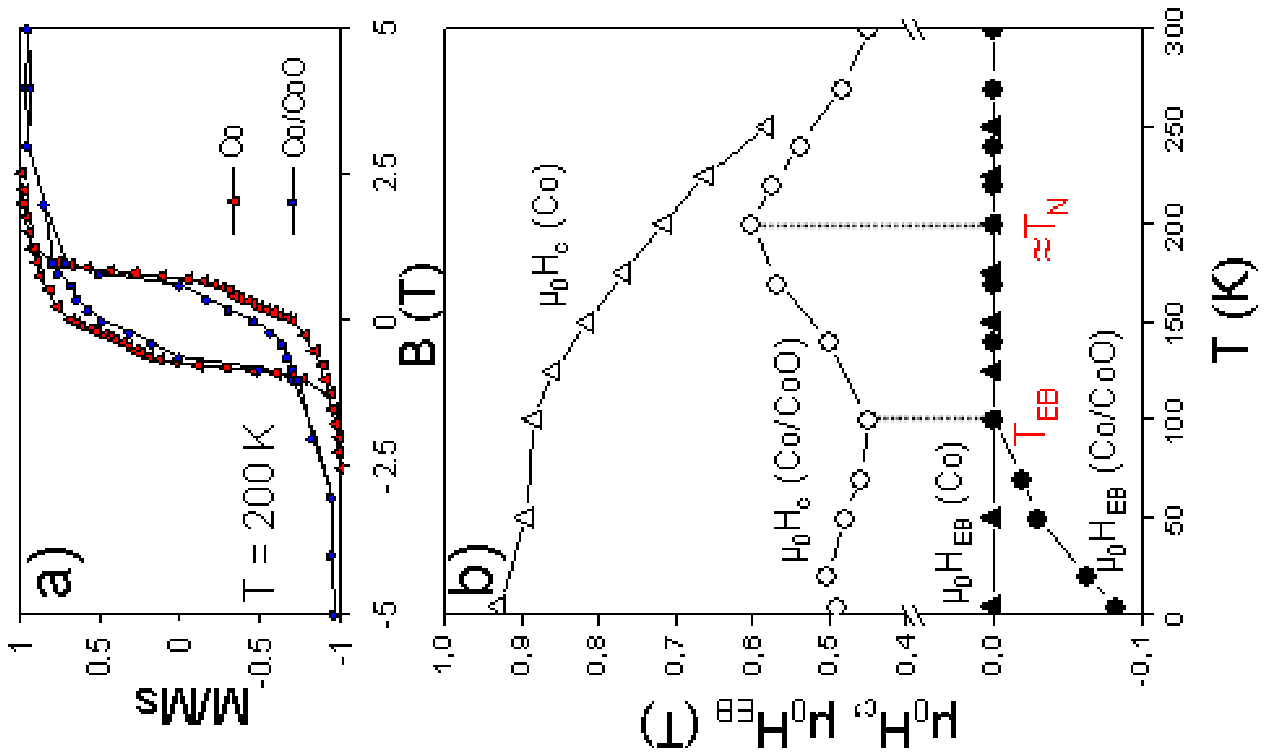} 

\caption{a) Typical hysteresis cycles obtained from the Co and Co/CoO nanowires.
b) Temperature dependence of the coercive field $\mu_{0}H_{C}$ (open
symbols) and the exchange bias field $\mu_{0}H_{EB}$ (filled symbols)
for non oxidized Co (triangles) and oxidized Co (circles) nanowires.
The samples were cooled under $B=5$ T and the measurements were performed
with an increasing temperature. The lines are guided for the eyes.}

\end{figure}

Below $T_{EB}$, the increase of $H_{C}$ with decreasing temperature,
along with the increase of $H_{EB}$, is in qualitative agreement
with previous studies \cite{berkowitz1999,nogues1999,peng2000,luna2004,iglesias2008,gruyters2000}.
However, the fact that the coercivity goes up upon warming between
$T_{EB}$ and $T_{N}$ is unexpected. As we shall argue, this is due
to the presence of superparamagnetic fluctuations of the AFM CoO grains.
For elongated systems, the main contributions to the coercivity of
the system arise from the shape anisotropy $K_{sh}$ of the wires
which is almost temperature independent and from the Co uniaxial magnetocrystalline
anisotropy $K_{mc}$ which decreases from $8\times10^{5}$ J/m$^{3}$
at $5$ K down to zero at $500$ K \cite{ono1980}. Thus a monotonous
variation of the coercive field would be expected. We argue that our
measurements unambiguously show that the temperature behavior of the
coercive field is related to the Exchange Bias (EB) phenomenon. The
comparison of the measurements on non-oxidized and oxidized Co wires
(see Fig. 4b) shows that the oxidation, and thus the EB mechanism,
leads to a drop of the coercivity of about $0.2$ T at the blocking
temperature.

To our knowledge, all reports on the Co/CoO system in the litterature
show that the coercive field monotonously increases below $T_{EB}$
\cite{peng2000,luna2004}. In a few reports, on some other exchange
bias systems, a maximum of the coercive field around the onset of
the blocking temperature $T_{EB}$ is observed (\cite{nogues1999}
and references therein, \cite{nishioka1998,eftaxias05-1}). We should
point out that these observations have been made on very low coercivity
systems where $H_{C}$ increase is only of a few mT and is attributed
to the increase of the AFM anisotropy around $T_{EB}$. In the present
case the effect is in the \emph{opposite direction} since we observe
a \emph{coercivity minimum} at $T_{EB}$. As evidenced by the temperature
dependence of $H_{C}$, the AFM surface layer modifies the core FM
magnetization up to almost $T_{N}$, which is well above the onset
of a static exchange bias at $T_{EB}$. Magnetization relaxation measurements
have thus been carried out to assert whether the observed exchange
bias effects is concomitant with a slowing down of the superparamagnetic
fluctuations of the AFM grains at the nanowire surface.

The magnetization relaxation was measured using SQUID (in a permanent
mode with 15 s extraction time) at small positive fields ($3$ mT)
after saturation under $5$ T (see Fig. 5a). The time decay of the
magnetization was first fitted using a phenomenological stretched
exponential expression: $M(t)=M_{c}+M_{0}\exp\left(-t/\tau\right)^{\beta}$,
where $M_{c}$ is the magnetization at infinitely long times (static
part), $M_{0}$ is the magnetic moment of the fluctuating volume,
$\tau$ is the relaxation time and $\beta=0.4$ is a stretch factor,
indicative of a distribution of relaxation times in the sample, giving
the best agreement with the experimental data . For single-size particles,
we would have $\beta=1$. The fact the $\beta$ is far off unity is
strongly indicative of a broad size distribution. For monodispersed
superparamagnetic objects with uniaxial anisotropy K, the temperature
dependence of the relaxation time $\tau$ is related to the energy
barrier $\triangle E$ separating the two stable states through the
Arrhenius expression: $\tau=\tau_{0}\exp\left(\triangle E/k_{B}T\right)$
where the energy barrier is driven by the total anisotropy energy
$K$ and the volume $V$ of the particles: $\triangle E=KV$. We define
$T_{N}=200$ K, the temperature at which the coercive field is maximum.
As shown in Fig. 5b, the relaxation time $\tau$ is first very short
above $T_{N}$ and then increases quickly upon decreasing temperature
down to 50K where it finally levels off down to the lowest temperature.
The behavior of the relaxation time, characterized by a progressive
slowing down of the relaxation, and the broad temperature range between
$T_{N}$ and $T_{EB}$ suggests that the CoO layer is composed of
a collection of anisotropic AFM grains with a broad size distribution
which will relax with a characteristic time controlled by their respective
energy barriers $\triangle E$.

\begin{figure}
\includegraphics[bb=30bp 50bp 530bp 500bp,clip,width=14cm,angle=270]{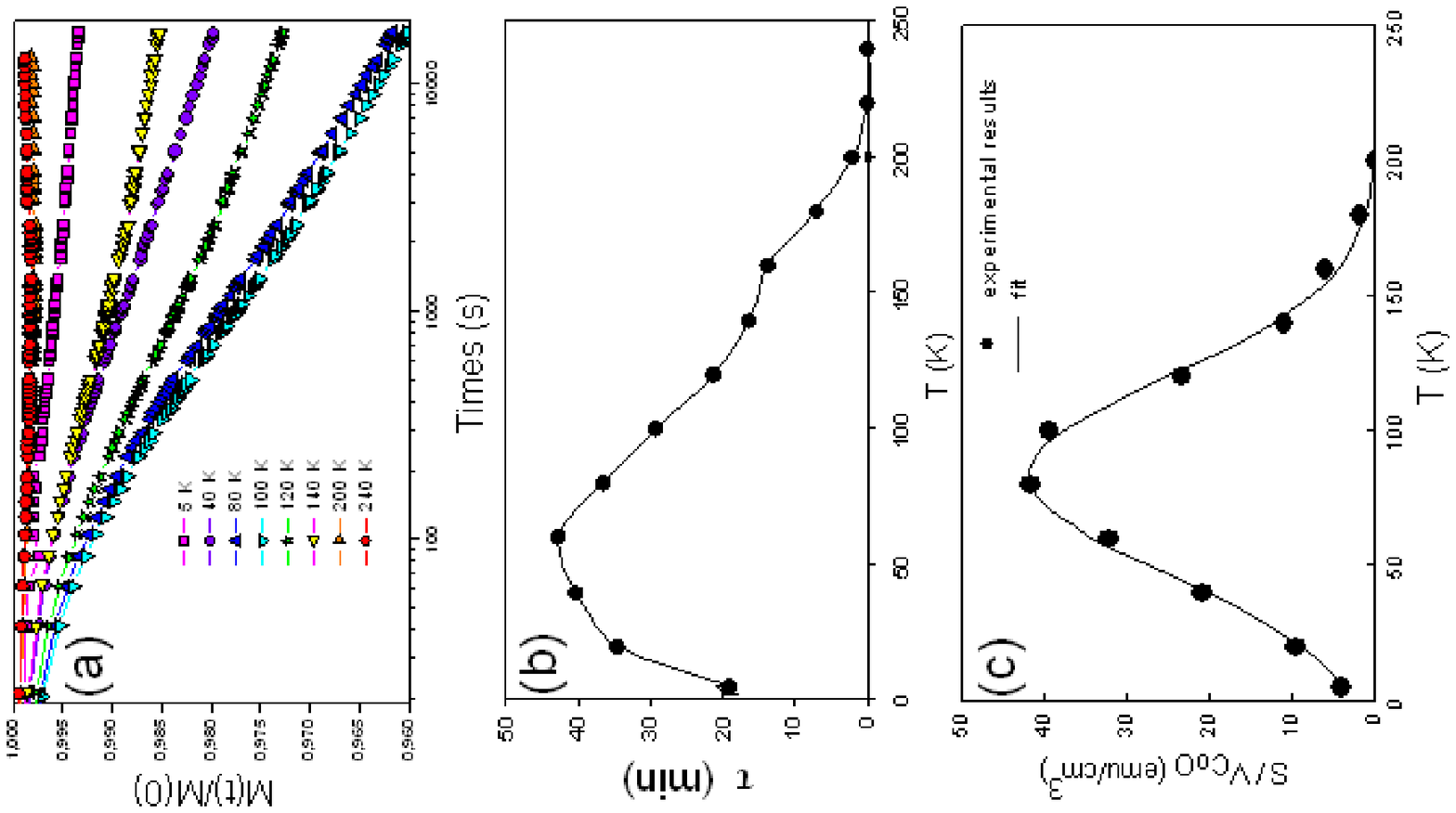}

\caption{(a) $M(t)/M(t=0)$ as a function of time for Co/CoO nanowires under
3mT after saturation at 5T. (b) Temperature dependence of the magnetization
relaxation time for oxidized Co/CoO nanowires extracted from the expression
$M(t)=M_{c}+M_{0}\exp\left(-t/\tau\right)^{\beta}$ with $\beta=0.4$
kept fixed throughout. (c) magnetic viscosity $S(T)$, as a function
of the temperature, extracted from the expression $M(t)=M_{0}-S(T)\: ln\left(t-t_{0}\right)${\large .}}

\end{figure}

In the case of a wide (almost flat) distribution of particle size
and anisotropy barriers, the behavior of the magnetization can be
described by the relation \cite{kneller1962}: $M(t)=M_{0}-S(T)\: ln\left(t-t_{0}\right)$,
where $S(T)$ is the magnetic viscosity. This dependence is well followed
in the time range $t>100$ s (see Fig. 5a). The viscosity parameter
is presented on Fig. 5c. At low temperatures, the viscosity $S(T)$
is low because most of grains are blocked and thus only a very small
fraction of the sample can relax. Upon, warming we observe a round
maximum at 80 K and then a steady decrease at higher temperatures.
The broad size distribution of the AFM grains means that, at a given
temperature, larger grains will tend to order along the FM magnetization
while smaller grains remain superparamagnetic. At low temperatures,
only the smallest grains will be superparamagnetic while the larger
ones are locked into one of their stable magnetization configuration;
hence a longer relaxation time in average and a smaller viscosity.
At high temperatures, the global viscosity of the system decreases
due to the fact only the few remaining large grains are contributing
to the relaxation \cite{pajic2007}.

The viscosity $S(T)$ is usually related to the distribution of energy
barriers through $S(T)=k_{B}TM_{S}/\triangle E{}_{m,T}$ where $M_{S}$
is the spontaneous magnetization of the CoO layer and $\triangle E_{m,T}$
is the mean energy barrier of the remaining grains that still relax
at a temperature $T$ \cite{Wohlfarth1984,Gaunt1986,StPierre2001}.
Larger grains, with higher energy barrier, are blocked while smaller
grains relax more rapidly than the time window of the measurement.
The quantity $\triangle E{}_{m,T}$ is equivalent to the inverse of
a distribution function $f(\triangle E_{m})$ (with $\int_{0}^{\infty}f(\triangle E_{m})d\triangle E_{m}=1$)
whose form can be either a flat distribution $(f(\triangle E_{m})=1/W$
between two extrema separated by $W$) or a Gaussian-like distribution
around a mean activated energy $k_{B}T^{*}$ : $f(\triangle E_{m})=A\exp((-k_{B}(T^{*}-T)/W)^{2})$
with $A=(1/\sqrt{\pi}W)(T^{*}/T)$. The best agreement is found for
the latter model with $T^{*}=83\pm1$ K and a width $W=35\pm2$ K
as shown in Fig. 5c. From the absolute values of the viscosity $S(T)$,
normalised by the volume fraction of CoO present in the nanowire (assuming
a 1.5 nm shell thickness), we find that the spontaneous magnetization
of the CoO shell is $M_{S}=15.2\pm0.1$ emu/cm$^{3}$, a value much
lower than the theoretical value ($224$ emu/cm$^{3}$). It implies
that the volume fraction which is {}``active'' represents only $7\%$
of the total volume of CoO in the materials. A similar result trend
was found in the case of granular CoO layers \cite{Gruyters2007}
or powders \cite{Dutta2008}. Equating the obtained mean activated
energies $k_{B}T^{*}$ for both compounds with the usual expression
for the energy barrier ($\triangle E=KV$) and assuming that the uniaxial
anisotropy is $K=5\times10^{5}$ J/m$^{3}$ \cite{gango1993,meiklejohn1957}
leads to active volumes of the CoO grains which are in the range of
$1.6-2.4$ nm$^{3}$.

To summarize, below $T_{N}$, the AFM moment fluctuations of the CoO
freeze progressively as the temperature is decreased \cite{Scarani};
leading to a low temperature rise of the relaxation time and a maximum
of viscosity below $T_{EB}$. Interestingly, we note that the static
part of the magnetization, $M_{C}$, is temperature dependent with
a sharp decrease below 60 K. This could be explained by the pinning
of the FM moments from the metallic core in contact with the AFM grains.
The physical origin of the superparamagnetism could be attributed
to a small fraction ($7\%$ as found from the experiment) of uncompensated
spins at the FM/AFM interface \cite{Roy2005,Tomou2006,Roy2007,inderhees2008,Blackburn2008}.

\section{Discussion}

The previous measurements give the following insight into the way
the exchange bias mechanism sets in our nanowires system. We have
unambiguously shown that the blocking temperature where the exchange
bias appears ($T_{EB}\approx100$ K) is well below the ordering temperature
of the AFM CoO shell ($T_{N}\sim220$ K). This can be accounted for
by the fact that the CoO shell is composed of small grains which are
subject to strong super-paramagnetic fluctuations down to rather low
temperatures.

We propose the following description of the magnetic properties of
our systems as a function of the temperature (see Fig. 6). Above the
Néel temperature the coercivity of the wires increases with decreasing
temperatures because of the increase of the magneto-crystalline anisotropy
of the Co. Below the Néel temperature $T_{N}$, a magnetic coupling
takes place between the CoO grains and the Co core of the wires, even
though all the CoO grains are still in a super-paramagnetic state.
In all reported systems such as spherical particles or thin films,
this leads to an increase of the coercivity by creating new loss mechanisms.
On the contrary, in our wires, we observe a significant drop of the
coercivity, by up to $0.25$ T, when the temperature is decreased.
The detailed mechanism of this coercivity drop is discussed in the
next section. The coercivity drop is larger with decreasing temperature
because: (i) the AFM moment increases when the temperature decreases
(see Fig. 3b), (ii) the AFM super-paramagnetic fluctuations slow down
(see Fig. 5b) which enhances the AFM-FM coupling. Eventually, at a
temperature $T_{EB}$, the largest CoO particles are blocked and this
gives rise to a finite exchange bias field $H_{EB}$. Below the temperature
$T_{EB}$, more and more CoO particles get blocked so that the exchange
bias field increases when the temperature is further decreased. Note
that the blocked CoO particle do not contribute anymore to the drop
of coercive field but only to the exchange field. Thus below $T_{EB}$,
the coercivity follows the same slope as the non oxidized wires (see
Fig. 4b).

The interactions between the AFM shell and the FM core directly reflects
in the magnetic viscosity temperature dependence. The magnetic viscosity
appears as soon as $T_{N}$ is reached. As the temperature is decreased,
and the AFM fluctuations slow down, the viscosity increases. Eventually,
below $T_{EB}$, as more and more CoO particles get blocked they do
not contribute anymore to the viscosity. At very low temperatures
where all the CoO particles are blocked, the viscosity becomes very
small.

We underline that this scenario is very different from the usual observations
in exchange bias systems. Experimental reports together with modelling
\cite{Fulcomer72-1,nishioka1998} show that when the temperature is
decreased, the coercive field $H_{c}$ increases to reach a maximum
at $T_{EB}$ and then $H_{c}$ decreases again when the temperature
is further decreased. In our system, because of the specific 1D geometry,
a minimum of coercivity is observed at $T_{EB}$. This is discussed
in the following section.

\begin{figure}
\includegraphics[bb=42bp 80bp 446bp 546bp,clip,angle=270,width=8cm]{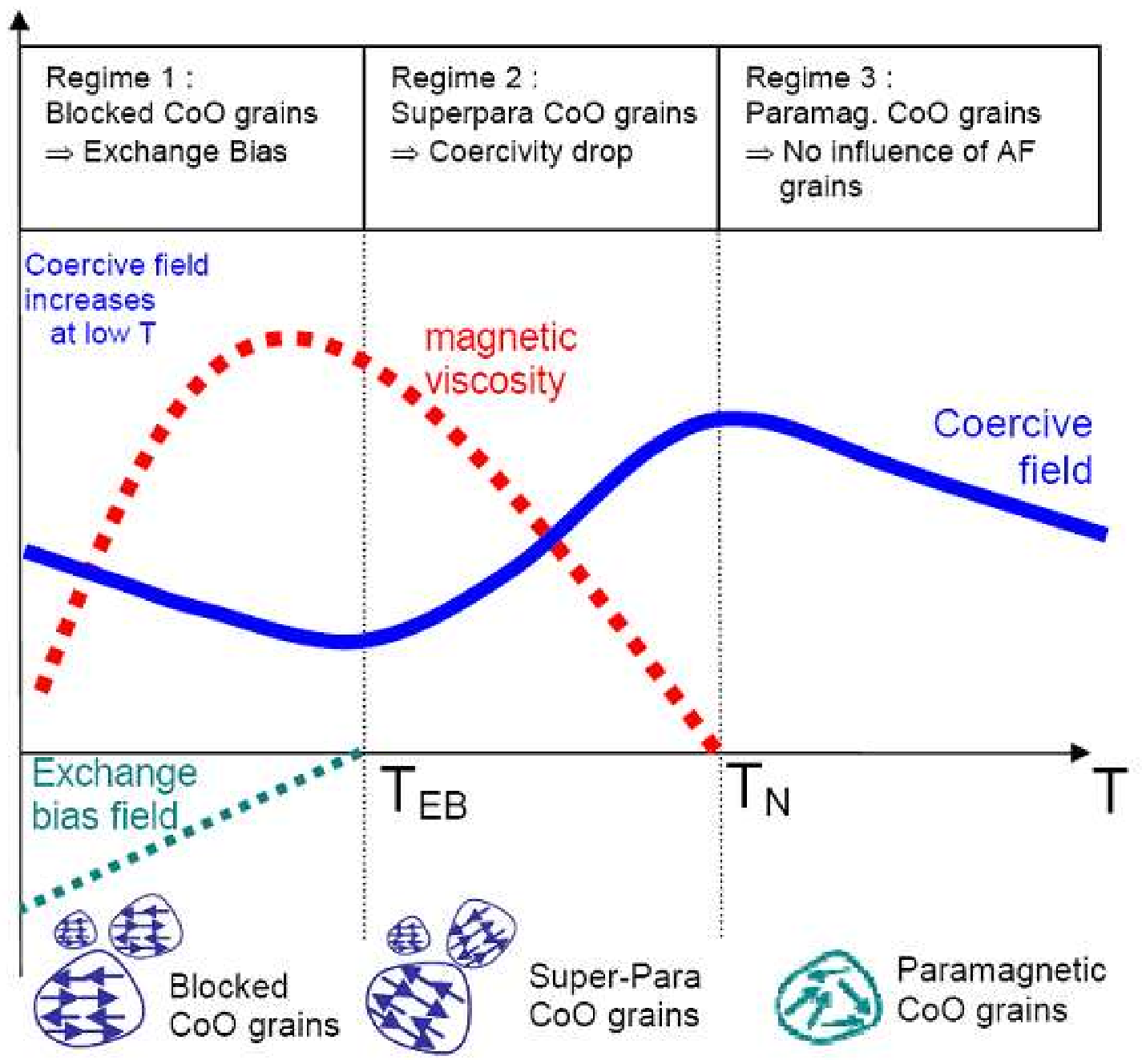}

\caption{Scenario of the magnetization and relaxation processes in oxidized
nanowires. We distinguish 3 regimes: (1) below the blocking temperature
$T<T_{EB}$, the CoO particles are blocked and a finite exchange bias
field appears; (2) between the blocking temperature and the Néel temperature
$T_{EB}<T<T_{N}$, the CoO particles are antiferromagnetically ordered
but are subject to superparamagnetic fluctuations; (3) above the Néel
temperature $T_{N}$, the CoO shell is not magnetically ordered and
there is no effective interaction between the FM wire core and the
AFM shell.}

\end{figure}

\section{Modelling}

As described above, the magnetic behavior of the nanowires is strongly
influenced by the oxide shell surrounding them. In order to qualitatively
understand the role of the interactions between a magnetic Co wire
and its CoO shell, we performed simulations with the \emph{Nmag} micromagnetic
modelling package \cite{fishbacher2007}. The studied model object
is a $100$ nm long, $10$ nm diameter cylindrical wire which is representative
of the experimental objects. The magnetic parameters used correspond
to typical values for hcp cobalt epitaxial thin films \cite{tannenwald1961},
saturation magnetization $M_{S}=1400$ kA.m$^{-1}$, exchange constant
$A=1.2\times10^{-11}$ J/m. The magnetocrystalline anisotropy is neglected.
The distance between two nodes of the mesh was taken four times smaller
than the exchange length $\ell_{ex}=\sqrt{2A/\mu_{0}M_{S}^{2}}\approx9.8$
nm (see Fig. 7).

As stated above, the key ingredients are the nanometer size CoO particles
which compose the shell around the wire. For the simulations, we consider
3 different regimes: (i) below the blocking temperature $T<T_{EB}$,
the CoO particles are blocked and a finite exchange bias field appears;
(ii) between the blocking temperature and the Néel temperature $T_{EB}<T<T_{N}$,
the CoO particles are antiferromagnetically ordered but are subject
to superparamagnetic fluctuations; (iii) above $T_{N}$, the CoO shell
is not magnetically ordered and there is no effective interaction
between the wire core and the shell.

\begin{figure}
\includegraphics[bb=60bp 100bp 350bp 650bp,clip,angle=270,width=8cm]{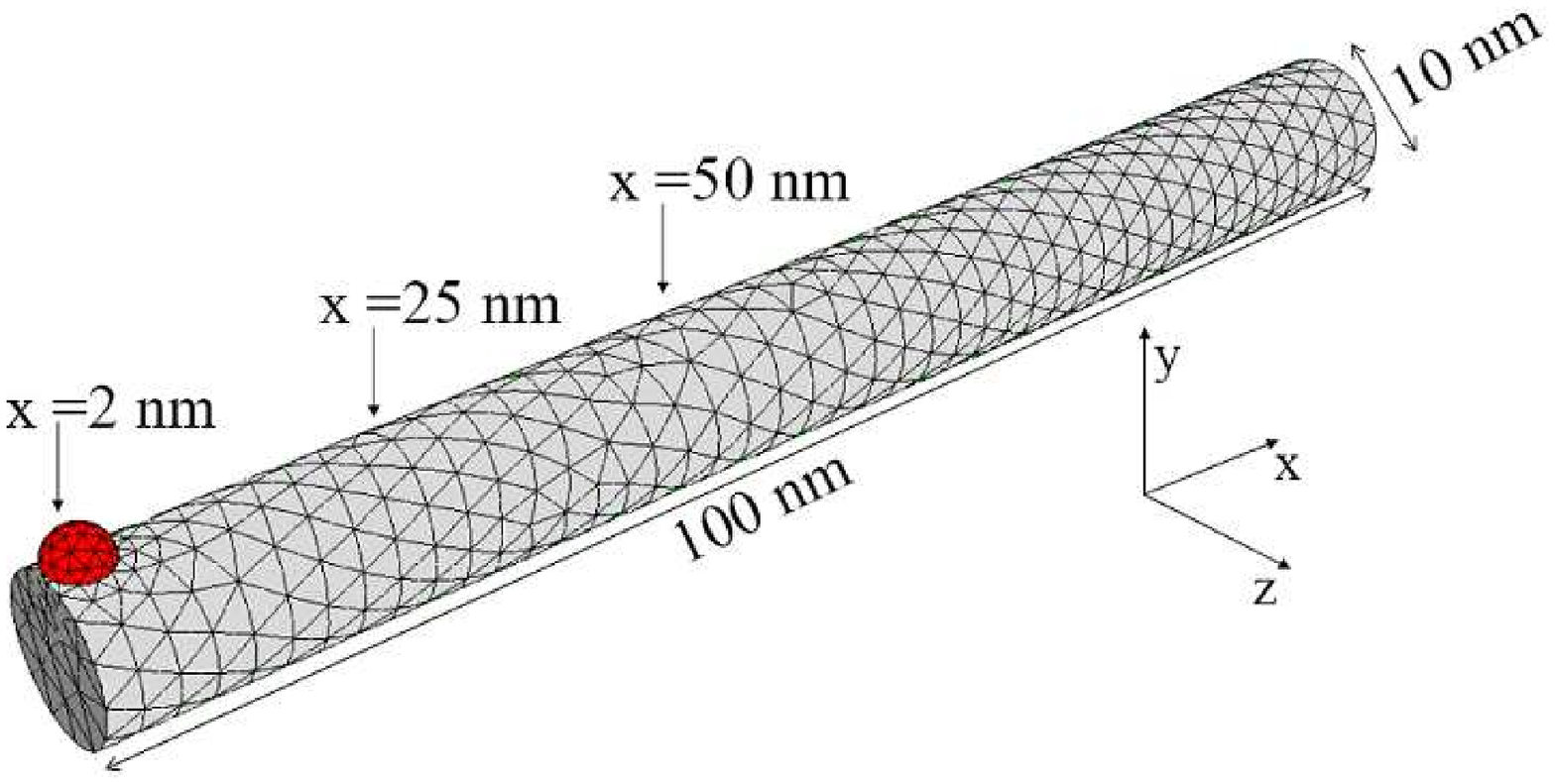}

\caption{Typical mesh used for the micromagnetic calculations in presence of
one hemisphere placed at the edge of the wire.}

\end{figure}

In the high temperature regime $T>T_{N}$, the simulation is straightforward
and lead to a coercive field of $471$ mT when the field is aligned
along the wire axis. Of course, in the case of randomly aligned wires
with respect to the field, the coercive field due to the shape anisotropy
is reduced by a few tens of mT due to the misalignement of the field
with the wires. When the temperature is decreased from room temperature
to $\sim200$ K, the coercive field increases as expected from the
magneto-crystalline anisotropy linear temperature dependance of Co
between $200$ and $300$ K \cite{ono1980} ($K_{mc}\approx5\times10^{5}$
J/m$^{3}$ at $300$ K and $\approx6.5\times10^{5}$ J/m$^{3}$ at
$200$ K).

In the low temperature regime, $T<T_{EB}$ , we consider that the
wire is coated with small particles, whose magnetic moments are blocked
along the x direction. These particles are modelled as half hemispheres
(see Fig. 7) and correspond to the blocked CoO particles. When the
temperature decreases the number of blocked AF particles increases.
Thus, in the simulations, we considered a wire coated with an increasing
number of such small blocked particles with a diameter of $4$ nm
(see Fig. 8 inset). Fig. 8 presents the evolution of the exchange
bias field $H_{EB}$ as a function of the total biased surface $S$
around the wire. The total surface of the wire is $3100$ nm$^{2}$.
One can observe that a few pinning points which represent only a small
fraction of the wire surface ($~7\%$) are sufficient to induce large
exchange bias fields ($\sim0.2$ T), which are of the same order of
magnitude as what is experimentally observed. Note however, that the
exchange at the interface was taken as $A_{FM-AFM}=1.2\times10^{-11}$
J/m which overestimates the efficiency of the exchange bias field.
Note also that the emergence of blocked grains barely affects the
coercive field. The coercive field is reduced from $470$ mT without
bias to $370$ mT with $7\%$ of biased surface. This qualitatively
explains why at low temperatures, when most of the AF grains are blocked,
the coercive field is not fully recovered in the oxidized wires compared
to the non oxidized wires (see Fig. 4b). It suggests that the $CoO$
grains act as nucleation points which promote the reversal of the
wires and reduce the coercivity.

\begin{figure}[!]

\begin{centering}
\includegraphics[bb=50bp 120bp 320bp 530bp,clip,angle=270,width=8cm]{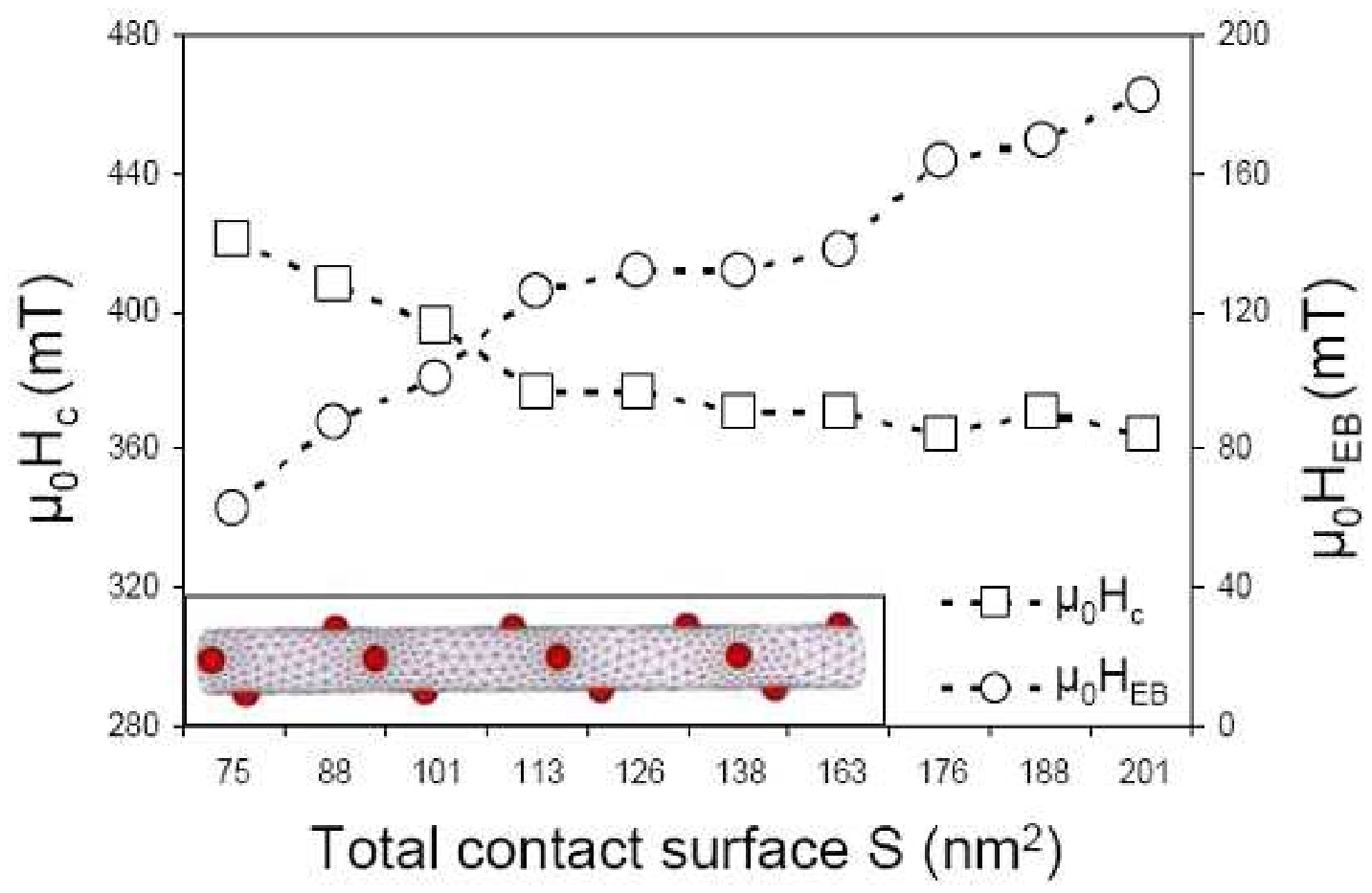} 
\par\end{centering}

\caption{\textit{\emph{$\mu_{0}H_{EB}$ and $\mu_{0}H_{C}$ measured on the
hysteresis cycle calculated along the wire axis in presence of more
and more blocked particles around the wire. $S$ is the total surface
in contact between the blocked particles and the wire. The particles
are homogeneously placed along the wire as shown in the inset.}}}

\end{figure}

In the intermediate regime, $T_{EB}<T<T_{N}$, the situation is more
complex. The relaxation measurements have shown that the CoO grains
in the shell have a broad size distribution range so that there is
also a broad distribution of the AFM fluctuation frequencies. It is
presently impossible or at least very difficult to tackle numerically
such a complex problem in the dynamic regime. Nevertheless, in this
intermediate regime, we think that is is possible to give some insight
of the role of fluctuating magnetic grains at the surface of the wires
provided some approximations are made. The first point to note is
that the characteristic reversal time of a $100$ nm Co wire is of
$4$ ns, as obtained from dynamic micromagnetic simulations using
a damping constant $\alpha=0.02$.

In the theory of superparamagnetism, the relaxation time $\tau$ is
related to the energy barrier $\triangle E$ separating two stable
states of a magnetic particle through the Arrhenius expression: $\tau=\tau_{0}\exp\left(\triangle E/k_{B}T\right)$.
The energy barrier is essentially driven by the uniaxial anisotropy
energy $K$ and the volume $V$ of the particles: $\triangle E=KV$
. The relaxation constant $\tau_{0}$ is of the order of $10^{-9}$
s. If we thus consider fluctuations of the AFM grains and use an anisotropy
constant $K=5\times10^{5}$ J/m$^{3}$ \cite{gango1993,meiklejohn1957},
the characteristic reversal time of $4$ ns corresponds to a volume
of the AFM particles of the order of $10$ nm$^{3}$. Smaller particles
will fluctuate much faster than the reversal time of the wire and
their interaction with the wire is likely to average out to zero.
Bigger particles will fluctuate much slower and can be considered
as static during the wire reversal. We thus make the assumption that
the very small $CoO$ grains will not play a key role in this intermediate
regime while the bigger particles will behave as static objects with
respect to the wire reversal so that static micromagnetic calculations
may provide realistic account of the interactions between the Co wire
and the CoO grains. The second assumption we are making is that the
CoO grains behave mostly as nucleation points for the magnetic reversal
of the wires. In order to model the CoO grains as nucleation points,
we modelled them as small ferromagnetic grains with their magnetization
free to rotate coupled to the Co wire with an exchange constant $A=1.2\times10^{-11}$
J/m. The CoO grains are thus represented as semi-hemispheres around
the Co wire (see Fig. 7).

We first assessed the role of the position of these nucleation points
along the Co wire (see Fig. 9a). The calculation was performed with
hemispheres of volume $17$ nm$^{3}$ ($2$ nm radius), the surface
$S$ in contact with the ferromagnetic wire thus being $12.5$ nm$^{2}$.
We find that the addition of such an hemisphere at the surface of
the wire can induce a significant drop $\delta H_{c}$ in the coercive
field ($\sim40$ mT) when it is placed close to the wire tip ($\delta H_{c}$
represents the difference of coercive field between the value obtained
from the isolated wire ($471$ mT) and the value obtained from the
wire surrounded by particles). On the other hand, such nucleation
points placed in the middle of the wire do not induce any drop in
the coercive field. The coercive drop can be almost doubled to ($\sim70$
mT) by simply putting a second symmetrical nucleation point. The sensitivity
to the nucleation point position can be explained by the distribution
of the demagnetizing field which is localized near the tips of the
wire and is close to zero in the rest of the wire \cite{ott2009}.
In the presence of an hemisphere located close to the tips of the
wire, the demagnetizing field interacts with the nucleation point.
This promotes an easier magnetization reversal and thus a smaller
coercive field. It is thus likely that it is mostly the CoO particles
located near the tips of the wires which are responsible for the coercivity
drop observed in our systems.

\begin{figure}
\includegraphics[bb=10bp 115bp 550bp 510bp,clip,angle=270,width=8cm]{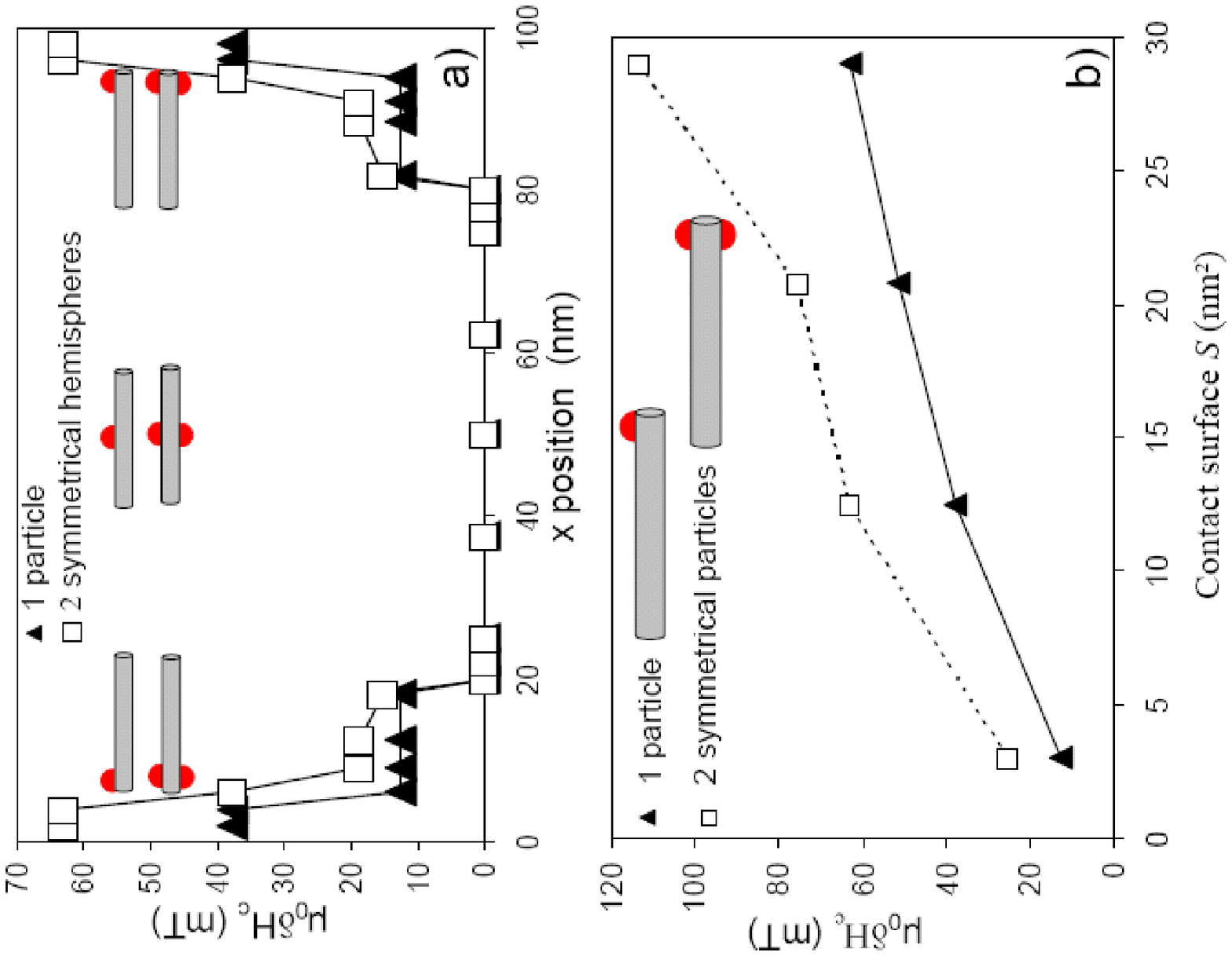}

\caption{(a) Drop of the coercivity ($\delta H_{c}=H_{c}(isolated\, wire)-H_{c}(wire+particles)$)
in presence of one (triangles) or two (squares) symmetrical hemispheres
of 2 nm radius at different positions along the lateral surface of
the wire. (b) Drop of the coercivity ($\delta H_{c}$) in presence
of one (squares) or two particles (triangles) of differents sizes
at the tip of the wire. $\delta H_{c}$ is proportionnal to the surface
in contact between the particles and the wire. Lines are guided for
the eyes.}

\end{figure}

We also investigated the effect of the nucleation point volume or
contact surface. Nucleation points of increasing contact surfaces
with the Co wire were considered (from $3$ to $28$ nm$^{2}$). Fig.
9b represents the drop in coercivity as a function of the contact
surface. It varies quasi-linearly from $12$ to $60$ mT for surfaces
$S$ varying from 3 to $30$ nm$^{2}$. The coercivity drop can be
doubled if two particles are placed symmetricaly at the end of the
wire. These calculated drops are of the same order of magnitude as
the ones experimentally observed. In the same way as before, large
grains placed far from the wires tips do not have any influence on
the coercive field. Contrary to the case of AFM grains at the surface
of a thin film, where an increase of the coercive field is usually
observed near the blocking temperature \cite{hoffman2003,nog2002},
we observe a drop of coercivity in our nanowires when the AFM grains
interact with the wire. This is due to the 1D geometry which is very
sensitive to the AFM grains which behave as nucleation points promoting
a magnetization reversal contrary to the case of thin films, where
AFM grains usually behave as pinning centers which drag the magnetization.

Fig. 10 presents three typical hysteresis in the three different temperature
regimes, for a magnetic field applied along the wire. The solid line
cycle corresponds to an isolated wire having no interaction with the
CoO particles ($\mu_{0}H_{c}\approx471$ mT). The long dash cycle
corresponds to a wire coated with nucleation points ($2\, nm$ radius
hemispheres) covering $7\%$ of the wire surface. The coercive field
is reduced to $360$ mT. The short dash cycle corresponds to a wire
coated with pinning points (2 nm radius hemispheres) covering 7\%
of the wire surface. The coercive field is still reduced to $360$
mT and a finite exchange bias field appears $\mu_{0}H_{EB}\approx145$
mT.

\begin{figure}
\includegraphics[bb=55bp 120bp 335bp 500bp,clip,angle=270,width=8cm]{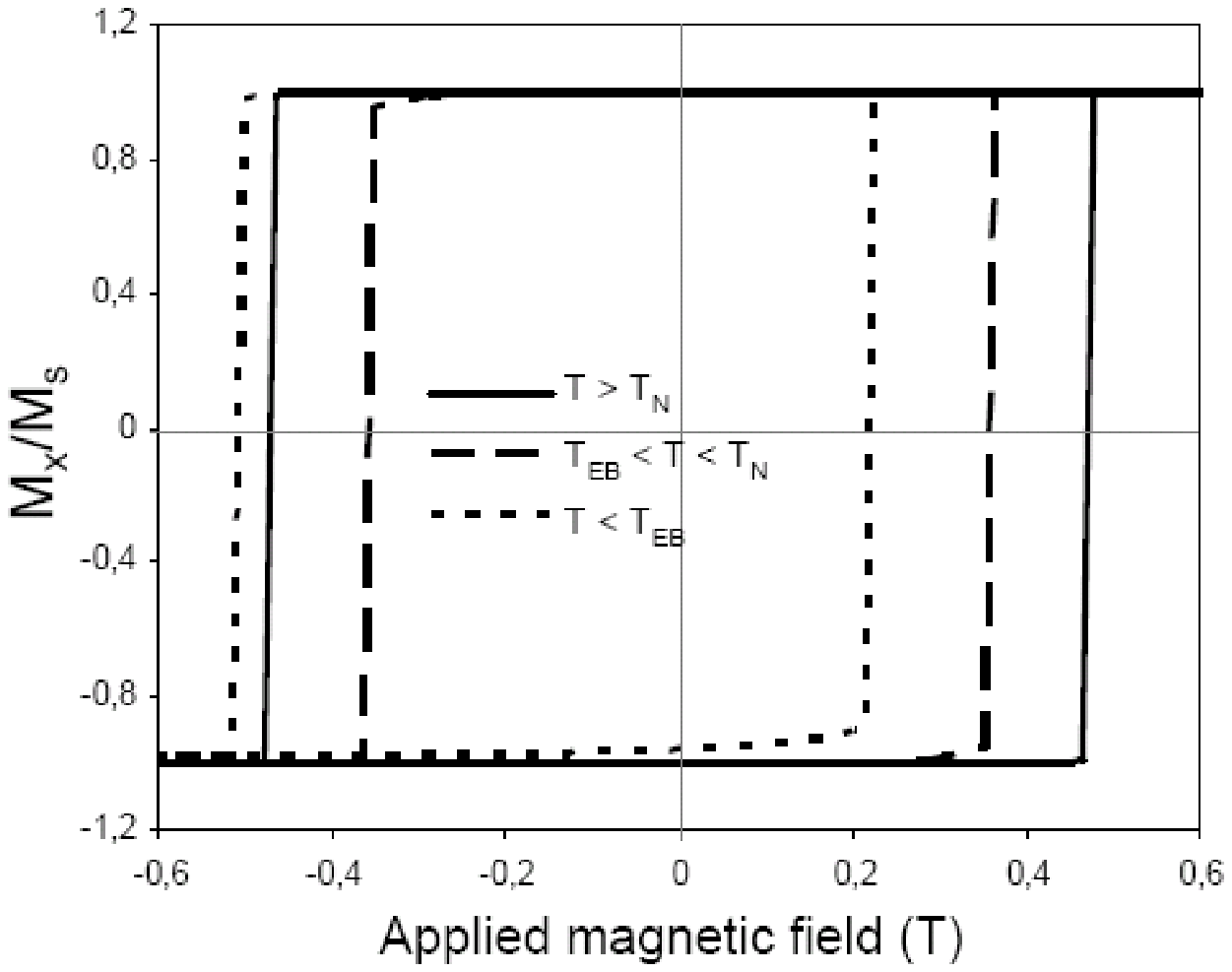}

\caption{Typical hyteresis cycles calculated \textit{\emph{in the three temperature
regimes. Solid line: $\mu_{0}H_{c}=471$}} mT \textit{\emph{, long
dashed line: $\mu_{0}H_{c}=360$}} mT \textit{\emph{, short dashed
line: $\mu_{0}H_{c}=365$}} mT \textit{\emph{and $\mu_{0}H_{EB}=145$}}
mT\textit{\emph{.}} }

\end{figure}

\section{Conclusion}

We have presented a study of the exchange bias phenomenon in Co/CoO
nanowires. We have shown that the AF ordering temperature of CoO oxidation
shell is rather high ($T_{N}\sim230$ K). The exchange bias field
reaches values of the order of $0.2$ T at low temperatures. We show
that a minimum of coercivity is observed around the blocking temperature
$T_{EB}\sim100$ K which is unambiguously related to the exchange
bias mechanism. Magnetization relaxation measurements show that this
effect finds its origin in the superparamagnetic fluctuations of the
oxidized AFM CoO layer. This proves that the exchange bias mechanism
sets in well above $T_{EB}$. Such a dramatic effect on the coercivity
properties was not observed in previous studies because 0D systems
(spheres) \cite{gango1993,skumryev2003,luna2004} and 2D systems (thin
films) \cite{vanDerZaag2000-1,nogues1999,Hou2000,Fulcomer72-1,devasahayam-1}
have a high degree of symmetry and low coercivities. On the other
hand, in the 1D geometry of nanowires the coercivity is dominated
by shape anisotropy effects. We suggest that the large drop of coercivity
is due to blocked AFM particles which act as nucleation points and
promote the magnetization reversal of the wires. This conclusion is
supported by micromagnetic simulations in which we can qualitatively
reproduce several of the features of the experimental measurements.
This study underlines the importance of the AFM super-paramagnetic
fluctuations in the exchange bias mechanism. 
\begin{acknowledgments}
The authors gratefully acknowledge the Agence Nationale de la Recherche
for their financial support (project P-Nano MAGAFIL). We thank F.
Herbst (ITODYS) for providing the TEM images of nanowires, M. Viret,
P. Bonville and J.B. Moussy (CEA-IRAMIS) for their help in the magnetometry
measurements and the \emph{Nmag} developers for their advices.\end{acknowledgments}


\begin{thebibliography}{52}
\bibitem{viau_rep}Non oxidized Co nanowires kept in 1,2 butane-diol
were transferred to 2,3 butane-diol adn placed in the SQUID cell for
magnetic measurements. 2,3 butane-diol has a melting point close to
room temperature allowing magnetic measurement opertaions on frozen
samples in the range 2-280 K.
\end{thebibliography}
\end{document}